\documentclass[reprint,superscriptaddress,aps,pre]{revtex4-2}

\usepackage{graphicx}
\usepackage{amssymb}
\usepackage{amsmath}
\usepackage{epsfig}
\usepackage{chemarr}
\usepackage{url}
\usepackage{natbib}
\usepackage{dcolumn}% Align table columns on decimal point
\usepackage{bm}% bold math
\usepackage{color}
\usepackage{braket}

\def\be{\begin{equation}}
\def\ee{\end{equation}}
\def\bmu{\begin{multline}}
\def\bea{\begin{eqnarray}}
\def\eea{\end{eqnarray}}

\begin{document}

%\title{Learning to fold}
%\title{Learning to fold at a bifurcation point}
%\title{Learning to reshape folding pathways}
%\title{Reshaping the topology of folding pathways through physical learning}
\title{Learning to self-fold at a bifurcation}
%\title{Reshaping the topology of mechanical bifurcations through physical learning}
\author{Chukwunonso Arinze$^1$, Menachem Stern$^2$, Sidney R. Nagel$^1$, Arvind Murugan$^1$}
%\title{Learning to self-fold at a bifurcation} %Reshaping the topology of folding pathways through physical learning}
%\title{Physically learning to fold}
%\title{Learning to fold along select folding pathways}
%\title{Learning to fold at a bifurcation}
%\title{Learning to fold at a bifurcation point}
%\title{Learning to fold along specific pathways}
%\title{Learning to chose folding pathways at a bifurcation}
%\title{Learning to self-fold at a bifurcation}

% Fig 4 - set up single and two vertex results?

\affiliation{Department of Physics, University of Chicago, Chicago, IL 60637}
\affiliation{Department of Physics and Astronomy, University of Pennsylvania, Philadelphia, PA 19104}

\begin{abstract}
%The folding of a creased thin sheet can be complex, with bifurcated folding pathways that branch out and reconnect in configuration space.
%A thin sheet with a crease network can fold in complex ways and faces choices, with bifurcated folding pathways that branch out and reconnect in configuration space.
%A thin sheet with disordered creases can fold in complex ways through a network of folding pathways that branch out and reconnect in configuration space.
%Metamaterial research has sought to rationally design non-linear deformation behaviors
Disordered mechanical systems can deform along a network of pathways that branch and recombine at special configurations called bifurcation points. Multiple pathways are accessible from these bifurcation points; consequently, computer-aided design algorithms have been sought to achieve a specific structure of pathways at bifurcations by rationally designing the geometry and material properties of these systems.
%Many attempts have sought to achieve specific pathway structure at bifurcations by rationally designing the geometry and material properties of these systems.
%Computer-aided optimization methods have sought rationally design the geometry of these
%Many attempts have sought to achieve specific pathway structure at bifurcations by rationally designing the geometry and material properties of these systems.
%to control these singularities of the energy function.
%structure of these systems
%A thin creased sheet can fold along a complex network of folding pathways that can branch out and recombine at bifurcation points in configuration space.
%A thin creased sheet can fold ..
%Metamaterial research has sought to rationally design these non-linear folding behaviors by designing the geometry and elastic properties of the creased sheet.
Here, we explore an alternative physical training framework in which the topology of folding pathways in a disordered sheet is changed in a desired manner due to changes in crease stiffnesses induced by prior folding. We study the quality and robustness of such training for different `learning rules', that is, different quantitative ways in which local strain changes the local folding stiffness. We experimentally demonstrate these ideas using sheets with epoxy-filled creases whose stiffnesses change due to folding before the epoxy sets. Our work shows how specific forms of plasticity in materials enable them to learn non-linear behaviors through their prior deformation history in a robust manner.
\end{abstract}

\maketitle

% Tasks:
% 5. Make Fig 2 into a full page
% 1. Theory SI
% 2. Experimental SI
% 3. Prep code for upload + coreldraw file
% 4. Fix figure refs.

%\section{Introduction}
% Design vs training
Metamaterials~\cite{liu2011metamaterials,reis2015designer} and smart materials~\cite{gandhi1992smart} are often designed to show specific behaviors. For example, mechanical topological insulators localize response to forces~\cite{bertoldi2017flexible,saremi2020topological} while elastic networks with allostery communicate deformations over a long range~\cite{rocks2017designing,tlusty2017physical,yan2017architecture}. More complex mechanical structures can show a specific deformation, e.g., a smile-shaped deformation, in response to a spatially textured pattern of forces~\cite{coulais2016combinatorial}. Most commonly, mechanical systems are rationally designed to show such behaviors by searching parameter space on a computer~\cite{diest2013numerical,jackson2019recent}. However, an alternate approach explored recently is that of \emph{physical learning}~\cite{pashine2021local,pashine2019directed,stern2020continual,stern2020supervised,stern2021supervised}: during a period of training, the material is shown examples of the desired behavior, prompting autonomous changes in the material parameters that promote the desired behavior. No computers are involved in optimizing the properties of such a system~\cite{dillavou2021demonstration,wycoff2022desynchronous,stern2021physical}.

% Open question and what has been done
Physical learning is a more constrained way of exploring parameter space than optimization on a computer. However, an autonomous physical learning process offers the advantage of learning from real examples of stimuli and response (as opposed to a theoretical specification of the problem) and does not rely on an accurate model of the material~\cite{wright2021deep}. Physical learning might also allow for continual learning of new functionalities \emph{in situ}, as requirements change~\cite{stern2020continual}. A major open question is what kinds of local adaptive processes available in real systems constrain physical learning~\cite{pashine2021local}. Recent work has shown that natural processes within an EVA foam network can train the network for an auxetic response by simply aging the material in different configurations~\cite{pashine2019directed}. Broader questions remain - what is the impact of different kinds of local learning dynamics on the feasibility and quality of learning? 

Here, we explore how the quality of physical learning depends on local adaptive processes (which we call `learning rules') through theoretical analyses and an experimental demonstration. We focus on training a fundamentally non-linear feature in marginal mechanical systems, a bifurcation point~\cite{kieffer1994differential,kumar2000computation,chen2011bifurcation,bigoni2012nonlinear,chen2014nonlinear,stern2017complexity}. Mechanical bifurcations occur at degenerate configurations from which multiple non-linear zero modes are accessible. Bifurcations cannot be described by a linear approximation even for small deformations since the energy vanishes to two leading orders; they correspond to singularities of the energy function with an excess of linearized zero modes and self-stress states that disappear at next to leading order.

Bifurcation points can potentially be exploited to create multi-functional systems~\cite{stern2020supervised} and have been studied in the context of mechanical linkages~\cite{Myszka2014-hx, Li2013-pk,rocklin2018folding} for robotics and topological meta-materials~\cite{chen2014nonlinear}. However, bifurcations can also make the system hard to control~\cite{kwatny2003static}. For example, folding outcomes at these singularities can be unpredictable and depend on the precise spatial pattern of forces used~\cite{stern2017complexity,chen2018branches,gillman2018truss,lee2021robust}. While generic mechanical systems show bifurcations in some parts of configuration space~\cite{wampler1986manipulator,wampler2011mechanism,Myszka2012-ap}, bifurcations are especially a problem for thin creased sheets because the flat state configuration is necessarily degenerate~\cite{stern2017complexity,chen2018branches}. In particular, creased thin sheets with nominally 1 degree of freedom (according to Maxwell counting), often called `self-folding origami', always have a bifurcation at the flat state that is the meeting point of an exponentially large number of distinct folding modes. Consequently, such `self-folding' sheets are hard to control at the flat state (despite the name) since the precise spatial pattern of forces applied will determine the folding geometry.
% Summarize what we do + what we find

We focus on the training of such creased thin sheets ~\cite{hull2002modelling,tachi2009generalization,dudte2016programming} to manipulate the bifurcation point in configuration space (Fig.\ref{fig:Vertex_schematic}A); in particular, we seek to increase the fraction of all force patterns that result in one folding pathway (i.e., attractor size, Fig.\ref{fig:Vertex_schematic}B) through a physical learning rule - crease softening due to folding (Fig.\ref{fig:Vertex_schematic}C, Fig.\ref{fig:Vertex_schematic}D). 
%We con, in which the stiffness of creases decreases with folding strain in that crease, 
We find that successful training relies on creating heterogeneity in stiffnesses across the sheet. However, this heterogeneity can rapidly diminish with further training for some classes of learning rules while other classes of learning rules, threshold-like in strain, are more robust. We test some of these ideas with an experimental prototype in which a sheet with epoxy-filled creases is folded back and forth along one pathway at a bifurcation. Such folding during the `training period' (i.e., before the epoxy sets) extrudes epoxy to different extents in different creases, creating a heterogeneous system. We test the trained sheet by applying different forces and find successful training in systems with 4 and 7 creases.

% pathway / mode
% positive or negative parts of a mode/branch/pathway

\section*{Results}

\subsection*{Training and Stiffness heterogeneity}

\begin{figure}
\centering
\includegraphics[width=0.45\textwidth]{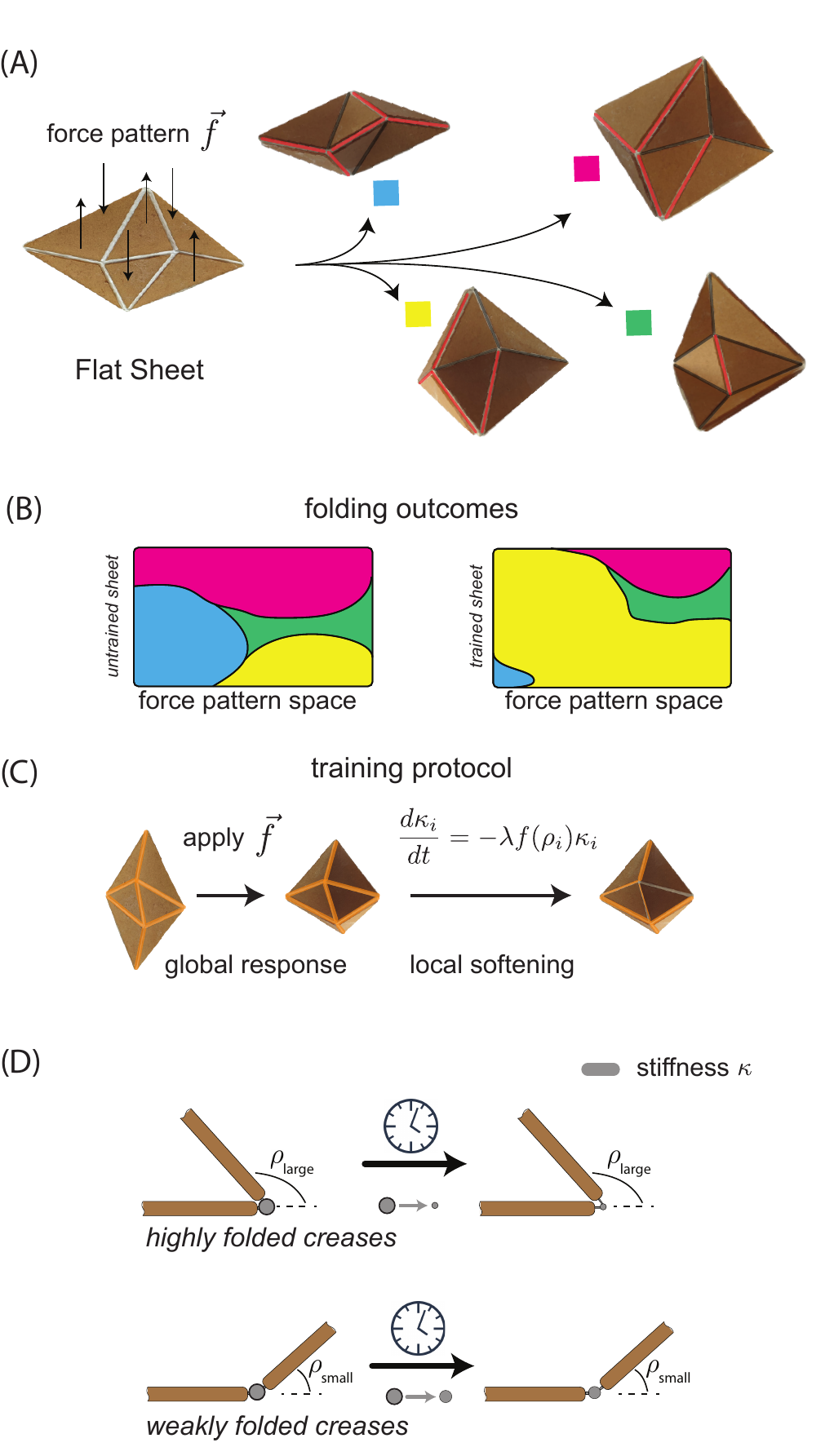}
\caption{The topology of folding pathways at a bifurcation can be changed through a physical training protocol.  (A) Thin creased sheets can fold along distinct folding pathways that bifurcate from the flat state with outcomes determined by the spatial force pattern. (B) A 2-d schematic representation of high-dimensional attractors in the space of force patterns. Each colored region represents one of the discrete folding outcomes that is realized for all force patterns represented by that region. Specific attractors can be enlarged, shrunk or eliminated by changing crease stiffnesses through design or physical learning. (C,D) We model a physical learning process that changes a crease $i$'s stiffness $\kappa_i$ (about the flat state) based on its folding strain $\rho_i$. } %Each folding mode has an attractor, composed of all spatial force patterns that result in folding along that pathway (schematic shown). The attractor structure can be changed by changing crease stiffnesses.} %Note that stiffnesses $\kappa$ about the flat state do not distinguish folding direction (mountain vs valley) at any one crease but nevertheless distinguish globally distinct folding modes. } % Why does this seem trivial? You fold along a pathway and now you can fold more.  % Not trivial - goes up and goes down. Not directional.
\label{fig:Vertex_schematic}
\end{figure}

% Bifurcation of sheets
We begin with a theoretical study of crease patterns made of 4-valent vertices as shown in Fig.\ref{fig:Vertex_schematic}A. Maxwell constraint counting gives this system 1 degree of freedom but in reality, this structure has two non-linear 1 degree of freedom pathways that meet at a bifurcation at the flat state~\cite{waitukaitis2015origami}. As a toy model of energy near a generic bifurcation, consider $E(x,y) = \lambda x^2 y^2$ where $x,y$ parameterize deformations of a mechanical system. Motions along $x=0$ and $y=0$ are true zero energy pathways, staying at zero energy for large deformations.  However, a linearized analysis at $(x,y)=(0,0)$ suggests a 2-dimensional vector space of zero modes (along with a self-stress state if we had a mechanistic model); further, any small deformation along $x=0$ or $y=0$ will reduce the zero mode space to 1 dimension (and eliminate the self-stress mode). Thus, any linearized analysis will fail to identify the true zero modes $x=0$ and $y=0$. On the other hand, any spatial pattern of forces applied to a 4-vertex~\cite{stern2017complexity} at the flat state bifurcation will result in folding along one of the two zero energy folding pathways.

%At these points, linearized analysis will predict a two or n-dimensional family of zero modes and (and accompanying self-stress states). But any deformation past this point collapses the system down to just one zero mode. These points arise generically in mechanical linkages \cite{ } and in sheets about the flat state. For sheets at the flat state, Maxwell count would say that there are so many zero modes. But in reality, there are even more zero modes at that point point , compensated by self-stress states to satisfy the generalized Maxwell count.
% design of attractor diagram
The non-linear force-response relationship of disordered mechanical systems at such bifurcations can be summarized by an `attractor diagram', shown schematically in Fig.\ref{fig:Vertex_schematic}B. The space is a 2-d schematic representation of high-dimensional space of spatial force patterns. Each colored region represents one of the discrete folding outcomes that is realized for all force patterns represented by that region. Earlier work has shown that this attractor structure can be changed by changing the geometry~\cite{stern2017complexity, chen2018branches}, pre-biasing preferred directions~\cite{tachi2017self,kang2019enabling, zhou2019biasing} of the sheet, or controlling the relative stiffness of different creases~\cite{stern2018shaping}.%; see Fig.\ref{fig:Vertex_schematic}C.

Typically, solving the inverse problem for attractors requires a complex computer algorithm; for example, Linear or Quadratic Programming algorithms on a computer~\cite{stern2018shaping} can determine crease stiffnesses that eliminate a chosen pathway in a saddle-node bifurcation. Here, we investigate whether the inverse problem can be solved by the sheet itself through a natural physical process, with no computers involved.

% Mountain valley symmetry
Throughout this paper, each crease, $i$, has a `crease stiffness', $\kappa_i$  which refers to the stiffness for the folding strain (or folding angle) about the flat state (i.e., unstrained/unfolded state) with rest angle maintained at zero; i.e., each crease, $i$ has a folding energy $(1/2) \kappa_i \rho_i^2$ for folding strain $\rho_i$ (see Fig.\ref{fig:Vertex_schematic}D) and crease stiffness $\kappa_i$. Thus, this stiffness does not prefer any folding orientation ($\rho \to -\rho$, which is known as mountain-valley symmetry) and only depends on the magnitude of the folding strain/angle of the creases, $\rho_i^2$. In contrast, a creased sheet of paper develops non-zero rest angles at each crease that breaks the mountain-valley symmetry ($\rho \to -\rho$) at creases and maintains the sheet away from the flat state bifurcation, without actually reshaping the bifurcation itself~\cite{stern2018shaping}. In this work, we focus on the non-trivial problem of shaping bifurcations, while maintaining mountain-valley symmetry at each crease, so the trained sheet can still be completely flattened. Consequently, our trained sheet can symmetrically access either the positive or negative components of a given folding pathway. % picking M or V does not count because that wouldn't allow a full range of motion and will prevent the sheet from being flat. )

\begin{figure}
\centering
\includegraphics[width=0.45\textwidth]{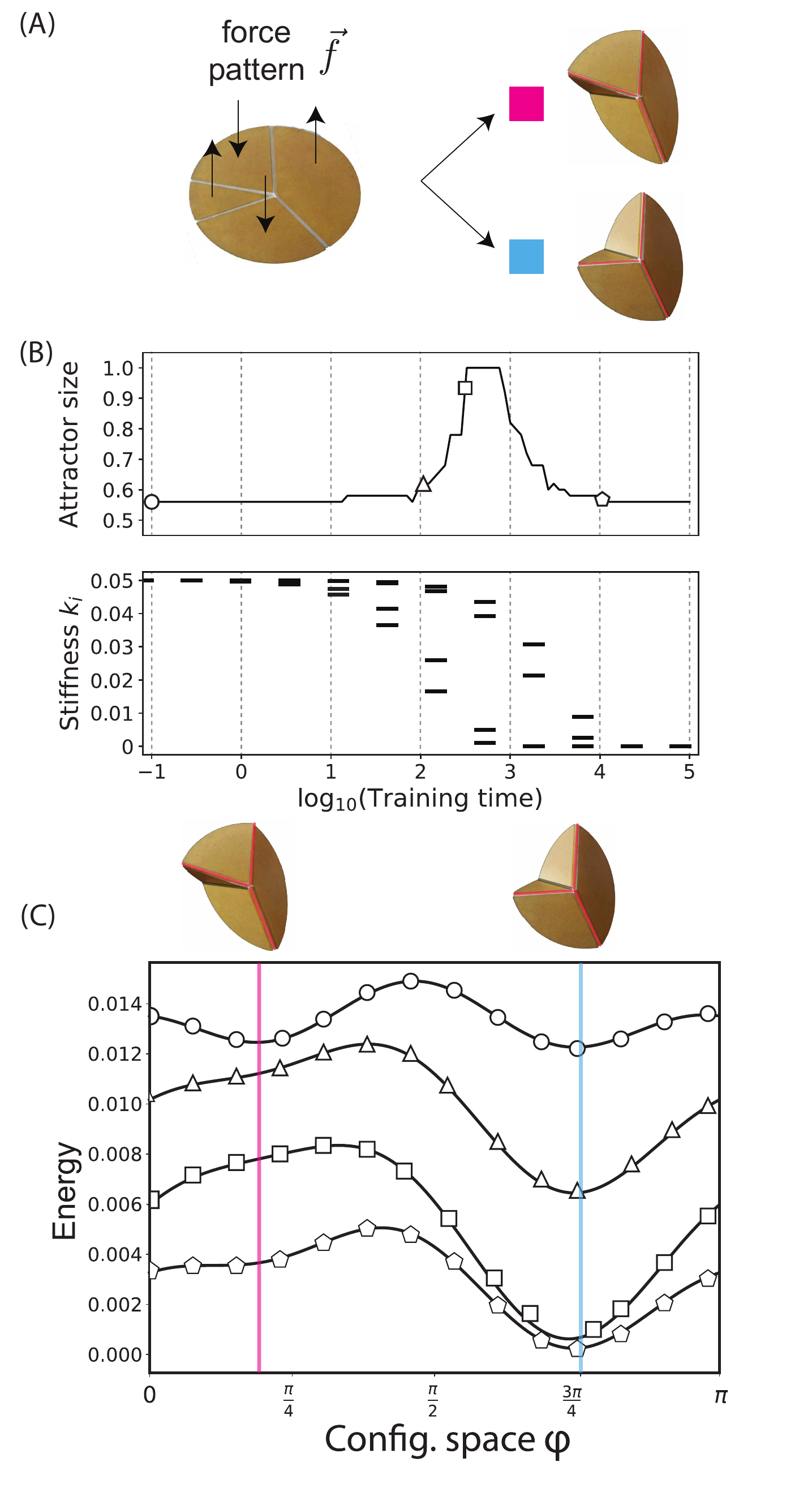}
\caption{Eliminating a select folding pathway for a 4-vertex through physical training. (A) A single 4-vertex has two distinct folding modes (blue, pink). (B) The blue pathway's attractor size during a training process in which the 4-vertex is folded by a fixed force pattern repeatedly. The stiffness $\kappa_i$ of each crease, $i$, are represented by four horizontal lines at each time point during training; all $\kappa_i$ start at the same stiffness and decrease according to Eq.\ref{eq:learning_rule}. $\kappa_i$ are in units of the bulk modulus of the stiff faces. Crease stiffnesses, $\kappa_i$, become heterogeneous during training, before becoming homogeneous again upon over-training. (C) Energy landscape of a sheet at different points during training ($\phi$ is an angular coordinate in 2-dimensional null space; see SI for details). Landscape before training (circles) and after over-training (pentagons) show two distinct minima; but when attractor size $\sim$ 1 (squares), pink minimum is eliminated via a saddle-node bifurcation. Note in the under-trained regime (triangle), the attractor size is below 1, but well above the initial value of 0.56} 
\label{fig:Vertex_training}
\end{figure}

As a first pass, we considered learning rules of the type shown in Fig.\ref{fig:Vertex_schematic}C and Fig.\ref{fig:Vertex_schematic}D that softens different creases based on the current folding strain:
\begin{equation} 
\label{eq:learning_rule}
    d \kappa_i/dt = -\lambda \rho_i^2 \kappa_i.
\end{equation} 
where $\lambda$, the `learning rate', sets the learning timescale. 

As a case study, we begin by applying the above learning rule to 4-vertex shown in Fig.\ref{fig:Vertex_training}A; the 4-vertex generically has two distinct folding branches that meet at the flat state bifurcation\cite{waitukaitis2015origami}. We simulated a training process in which the vertex initially has creases of equal stiffness $\kappa_i^0$; the sheet is folded repeatedly (using a spatial pattern of forces) along the positive and negative components of one of the two pathways at the bifurcation. The sheet is folded to a configuration of finite strain and the parameters $\kappa_i$ are updated according to learning rule above for a time interval $\tau$. The sheet is then relaxed back to the flat state and re-folded with the negative of the same pattern of forces and the learning process is continued. {See SI for parameters.} We test the attractor size of each pathway throughout training; the attractor size is determined by applying a library of $50$ randomly chosen spatial pattern of folding torques to the creases and counting the fraction of folding torques that result in the chosen folding pathway. We assume that testing does not cause further changes in the stiffnesses $\kappa_i$ in our simulations (though a real system would continue changing due to testing).

We find that during training, different creases fold to different extents $\rho_i$ in specific ratios characteristic of the branch we fold along. Consequently, the learning rule Eq.\ref{eq:learning_rule} creates heterogeneity in the initially homogeneous creases stiffness $\kappa_i$. When crease stiffnesses are relatively heterogeneous, the attractor size of the chosen pathway increases from an initial value of $0.56$ to nearly $1$, i.e., nearly all spatial patterns of forces result in the chosen folding pathway; see Fig.\ref{fig:Vertex_training}B. However, further training reduces the heterogeneity in creases as all stiffnesses approach zero. In this `overtrained' regime, the attractor size of the desired pathway drops down to $0.56$ again, no bigger than for the initial untrained sheet; see Fig.\ref{fig:Vertex_training}B.

To illustrate the phenomenon of saddle-node bifurcation, we computed the energy of the origami sheet in different folded configurations, while its crease stiffnesses,$\kappa_i$ were evolved by the learning rule, Eq.\ref{eq:learning_rule}. We studied folded configurations within the null space of the origami's potential at the flat state. This null space is a 2-dimensional space ~\cite{stern2018shaping}. We parameterize this null space with the variables $r$ and $\phi$. We selected a circle (defined as $r=0.5$) in this null space and computed the energies of various folded configurations represented by points on the circumference of this circle. An energy minimum in this energy plot (Fig.\ref{fig:Vertex_training}C)  corresponds to the existence of a stable folded configuration. For untrained sheets, we see two minima corresponding to the two true non-linear folding modes. During training, the attractor size of one of these true non-linear folding modes is reduced to zero, as the mode is destroyed in a saddle-node bifurcation. This is illustrated in Fig.\ref{fig:Vertex_training}C (see pink line), as the local minimum on the left gradually transforms to a local maximum in the course of training. In the overtrained regime, the eliminated mode (local minimum on the left) reappears and we find that both pathways are accessible in response to some spatial patterns of forces (See images of the two modes atop Fig.\ref{fig:Vertex_training}C).

%Conclusion:
Hence, we find that training can solve an inverse problem for non-linear behavior, namely that of eliminating one select branch in a saddle-node bifurcation and, thus, changing the topology of pathways. Further, successful training is correlated with the development of stiffness heterogeneity (see Fig.\ref{fig:Vertex_training}B); this observation is an example of a larger principle that disordered systems can learn because the information must be stored in the trained degrees of freedom; homogeneous creases cannot store such information.  Finally, we find that this particular training rule is prone to over-training and homogenization of creases if training is carried on for too long (see Fig.\ref{fig:Vertex_training}B).

We have found that heterogeneity in the crease stiffnesses stores the learned information about the desired pathway; but the learning rule that created the desired heterogeneity also erases that information upon further training. Similar erasure of trained response was observed in systems like cyclically sheared Brownian suspensions and charge-density wave conductors~\cite{keim2019memory}. %We leave a broader discussion of such information erasure phenomena to future work.

\begin{figure*}
\centering
\includegraphics[width=\textwidth]{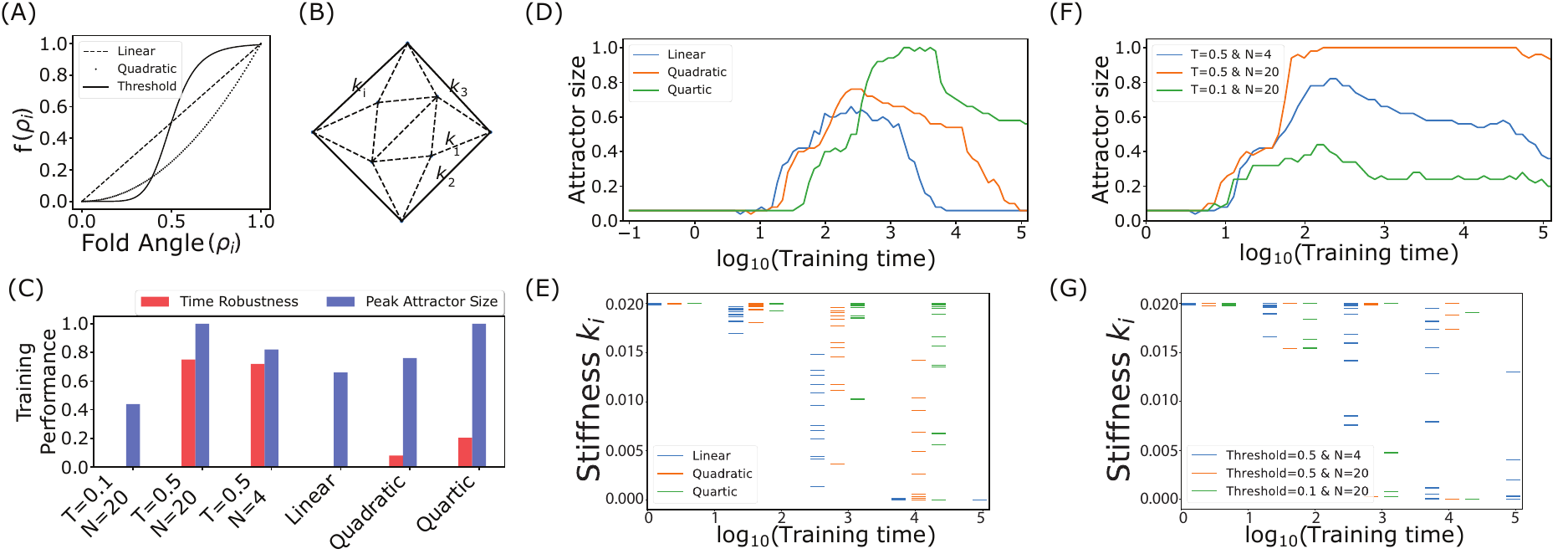}
\caption{Quality and robustness of learning depend on the learning rule. (A,B) For the sheet with 13 creases shown, we model a family of `training rules' with crease stiffness $\kappa_i$ that change with strain $ \rho_i$ (which has been normalized and, thus,ranges from 0 to 1) in a polynomial $f(\rho_i) \sim \rho_i^N$ or threshold $f(\rho_i) \sim \rho_i^M/(\rho_i^M+T^M)$ manner. (C) Learning quality (peak attractor size; see D,F) and robustness (length of training time over which attractor size $> 0.80$; see D,F) for different learning rules are explored here. 
(D) With  $f(\rho_i) \sim \rho_i^N$ rules, the attractor size of the desired pathway rises to a peak before falling. The peak attractor size (quality of learning) increases with $N$. (E) The stiffness of each of the 13 creases, $i$, of the origami are represented by the 13 horizontal lines. They all start out with the same stiffness and decrease according to the different polynomial learning rules indicated by the colors of the line. $\kappa_i$ are in units of the bulk modulus of the stiff faces. The polynomial rules also show a regime of over-training where stiffness $\kappa_i$ become homogeneous again as all $\kappa_i$ go to zero. The horizontal lines representing the crease stiffnesses for the different polynomial learning rules are plotted at select times, so as to avoid the lines of the different polynomial learning rules from overlapping. (F) Attractor size vs training time for threshold learning rules. (G) The stiffness of each of the 13 creases, $i$, of the origami are represented by the 13 horizontal lines. They all start out with the same stiffness and decrease according to the different threshold learning rules indicated by the colors of the line. $\kappa_i$ are in units of the bulk modulus of the stiff faces. Crease heterogeneity grows with training, and does not decrease (no overtraining) with continued training for some of the threshold learning rules (orange line).The horizontal lines representing the crease stiffnesses for the different threshold learning rules are plotted at select times, so as to avoid the lines of the different threshold learning rules from overlapping. %learning is robust to overtraining for sigmoidal learning rules. %However, learning is sensitive to choice of threshold $T$. 
 }
\label{fig:learning_rules}
\end{figure*}

\subsection*{Robustness of different learning rules}

Qualitatively, many real materials soften with strain as captured by the learning rule Eq.\ref{eq:learning_rule}, but might differ quantitatively. Different materials will have different learning rules; some might soften proportionally to their strain or to higher powers of their strain and yet others might be even more sensitive, softening only for strain above a specific threshold.  Other non-equilibrium systems can show more complex learning, where synaptic weights or learning degrees of freedom can both strengthen or soften over time; we do not investigate those cases.

%We sought to understand whether such material properties could impact the quality of learning and if some choices are preferred.
We investigated whether such quantitative differences in material properties might have a qualitative effect on the quality and robustness of learning. We considered families of rules of the type:
\begin{equation} \label{eq:broader_rules}
    d \kappa_i/dt = -\lambda f(\rho_i) \kappa_i.
\end{equation} 
The first family we considered were different polynomial forms $f(\rho) \sim \rho^N$. Note that Eq.\ref{eq:learning_rule} is the case where N=2  (See Fig.\ref{fig:learning_rules}A). The second family, defined by a Hill coefficient $M$, $f(\rho) \sim \rho^M/(\rho^M + T^M)$ corresponds to sigmoidal dependence often seen in real systems; small strains do not cause significant aging or change in stiffness but strains above a characteristic threshold $T$ cause stiffness changes; further the precise amount of strain does not matter beyond this threshold $T$ (See Fig.\ref{fig:learning_rules}A).

We trained a larger disordered creased sheet with 13 creases and 4 vertices (See Fig.\ref{fig:learning_rules}B); this sheet had 16 non-linear modes meeting at the flat state bifurcation. We used the same training protocol as for the single vertex in Fig.\ref{fig:Vertex_training}: we selected one pathway as the desired pathway and applied the learning rule as the structure was folded repeatedly into the positive and negative components of the selected pathway. As earlier, we quantified successful learning by the attractor size of the desired pathway, i.e., folding the sheet with a library of $50$ random force patterns and computing the fraction of force patterns that result in a specific folded mode. We assume that stiffnesses do not change during such testing.
%We found that with polynomial $f(\rho) \sim \rho^n$, the higher order polynomials (higher $n$) were better at creating heterogeneity that gave successful learning over a greater range of training times.  

Among polynomial learning rules $f(\rho) \sim \rho^N$, we found that training rules  with higher order polynomials (higher $N$) resulted in better training. We found a higher peak attractor size and training was also more robust - learning quality stayed higher for longer (See Fig.\ref{fig:learning_rules}C). However, all polynomial learning rules were still susceptible to overtraining, during which the crease stiffness heterogeneity was lost (See Fig.\ref{fig:learning_rules}D, Fig.\ref{fig:learning_rules}E). 
With threshold-like learning rules, the over-training problem was nearly eliminated (See Fig.\ref{fig:learning_rules}F). We found that creases that fold less than the threshold $T$ do not soften at all and hence the learned heterogeneity in stiffness is maintained over time. However, there is a trade-off; the threshold $T$ of the learning rule needs to be within the range of strains experienced during training. If $T$ is too large, no training would occur. If $T$ is too small, training will fail to create sufficient heterogeneity in stiffnesses $\kappa_i$ to encode information about the desired mode (See Fig.\ref{fig:learning_rules}G). 

\subsection*{Experimental demonstration} 

With these theoretical results in place, we demonstrated these ideas with an experimental prototype. While many previous works have implemented creased sheets in systems ranging from graphene on the nanoscale~\cite{miskin2018graphene} to mylar and cardboard on the mesoscale~\cite{pinson2017self} to solar panels on satellites~\cite{miura1985method}, these works generally have fixed stiffness in different creases and thus an inability to learn folding behaviors. We note that paper or cardboard is affected by folding but typically develops a non-zero rest angle upon folding and is thus likely to fold in the same way again. But as noted earlier, such a displacement from the bifurcation does not reshape the bifurcation which still exists if the sheet is forced into the flat state. Here, we create a prototype that maintains $\rho \to - \rho$ symmetry at each crease and hence can still be laid flat after training. % we focus on stiffness at the flat state with no rest angle.

\begin{figure}
\centering
\includegraphics[width=0.5\textwidth]{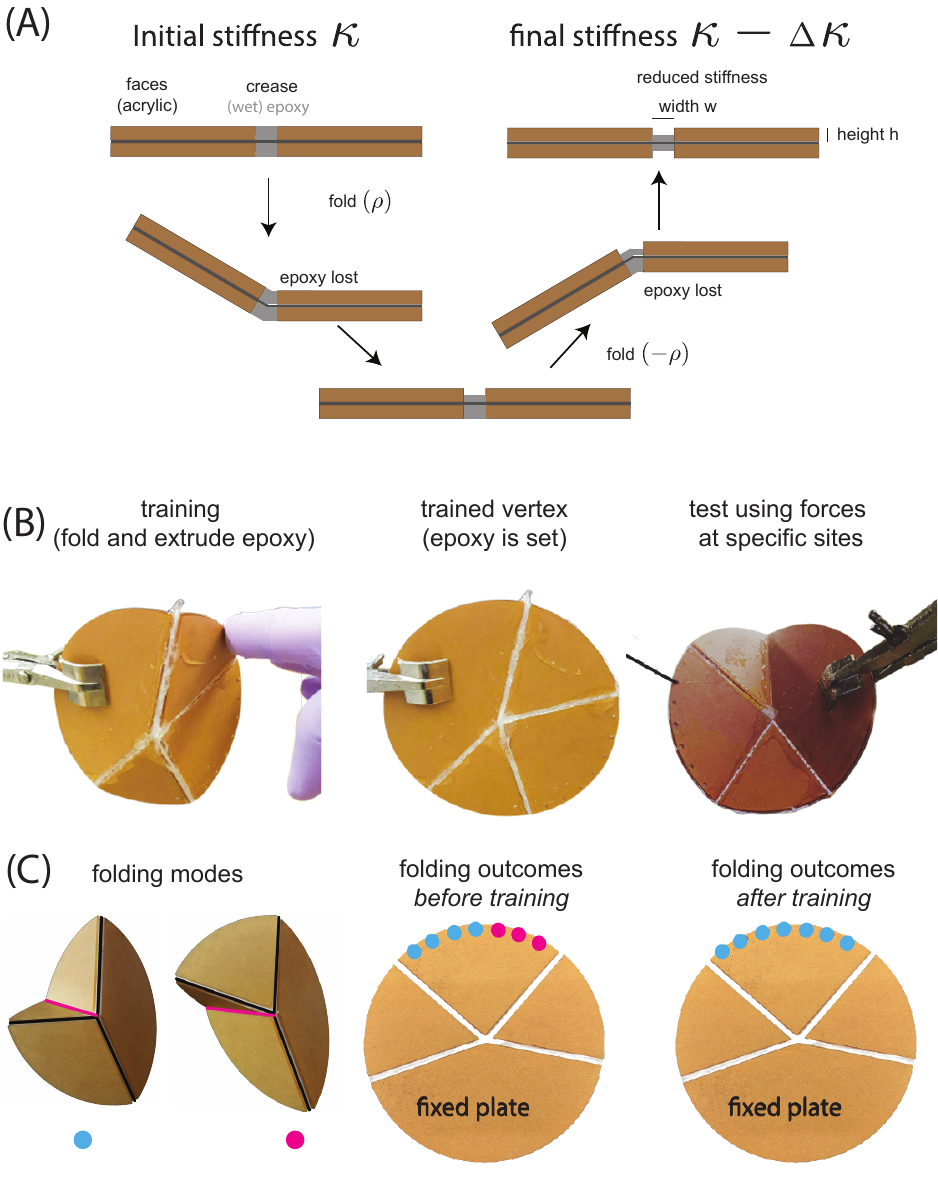}
\caption{Experimental realization of training through epoxy extrusion. (A) (schematic) We created creased sheets by sandwiching a thin membrane between thick acrylic sheets that serve as stiff faces; the resulting creases of width $w$ and height $h$ are filled with epoxy (orange). Folding by angle $\rho$ before the epoxy sets (the training period) will extrude epoxy; creases with larger $\rho$ will extrude more epoxy, resulting in lower stiffness $\kappa$ after the epoxy sets.
(B) A 4-crease vertex with epoxy filled creases was trained by repeatedly folding along one of the two pathways during the setting period; folding is repeated along positive and negative components of the chosen pathway to avoid any directional folding bias in the creases. (C) Testing: Folding outcomes are determined for folding forces applied at different locations. The untrained sheet folds along two distinct pathways depending on location of applied force (pink and blue dots). The trained sheet folds along only one pathway for all folding forces (blue dots), demonstrating an increased attractor for the blue pathway. 
}
%(B) We experimentally characterized the learning rule $f(\rho)$, i.e., the relationship between folded strain $\rho$ and final stiffness $\kappa$. Samples were folded to different strains $\rho$ and flattened before the epoxy set; the resulting final crease stiffness $\kappa_i$ was measured.}% For small crease widths, we find a strain-stiffness relationship $\Delta \kappa_i \sim \rho$ with $n = 4$.
\label{fig:experiment_setup}
\end{figure}

We created a sheet with gullies at creases by sandwiching a Tyvek sheet between acrylic pieces that were laser cut to serve as stiff faces of a crease pattern. See Fig.\ref{fig:experiment_setup}A. Consequently, the creases correspond to gullies of width $w$ (set by the gap between adjacent acrylic pieces) and depth $h$ (set by the thickness of acrylic) on both sides of the sheet. A slow setting epoxy solution is created from a mixture of epoxy resin and a curing agent in the ratio $1:2$. The creases are filled with epoxy on both sides of the sheet; the epoxy takes $\sim 60$ minutes to set. See SI for details. During this setting time (the `training phase'), the crease is folded to an angle $+\rho$, flattened again and folded to angle $-\rho$ in the other direction. Such folding will extrude an amount of epoxy $h(\rho)$ from the crease gully that depends on the magnitude of folding $\vert\rho\vert$. Only epoxy remaining inside the crease determines the stiffness of that crease at the end of the training phase. Thus, the amount of epoxy extruded $h(\rho)$ determines the change in stiffness $\Delta \kappa$ during training and thus determines the form of the learning rule in Eq.\ref{eq:broader_rules}. By folding in both directions ($\pm \rho$) during training, we maintain mountain-valley symmetry and zero rest angles at the crease; consequently, the trained sheet can still be laid flat.The training protocol is illustrated in Fig.\ref{fig:experiment_setup}B.

%Exactly how much epoxy is extruded by a crease is response to a strain $\rho$ presumably depends on the geometry of the crease.
We constructed a vertex with 4 creases (studied theoretically in Fig.\ref{fig:Vertex_schematic}), resulting in two folding modes (pink and blue) shown in Fig.\ref{fig:experiment_setup}C. Initially, before any epoxy is added, all creases are free folding. We tested the response of this free folding vertex to forces applied at different points along the boundary of one of the sectors. We found that forces at 3 of 7 locations leads to the blue folding mode while forces at the 4 locations result in the pink mode (See Fig.\ref{fig:experiment_setup}C).  Forces applied to all points in the other sectors led to either the blue or the pink mode. We then filled the creases with slow setting epoxy, marking the start of the training phase. We folded the vertex into a selected mode (blue) with characteristic folding angles $\rho_i^{blue}$ at each crease, reverted to the flat state and folded along the negative component $-\rho_i^{blue}$ of the same blue pathway. We folded until the largest folded crease could not be folded further; in this way, the magnitude of folding is approximately the same along the positive and negative components of the pathway and over multiple instances of folding during training. Throughout this training, the vertex was clamped in a vertical configuration but was rotated periodically to prevent epoxy from flowing out of the creases due to gravity. We repeated this folding process for 60 minutes, folding back and forth along the positive and negative components of the blue mode.  See SI for details of the experiment.

After the epoxy set, we found that forces at all 7 test locations now lead to the blue mode as illustrated in Fig.\ref{fig:experiment_setup}C, showing that the training procedure had modified the flat state bifurcation, presumably by eliminating the pink branch at a saddle-node bifurcation away from the flat state~\cite{stern2018shaping}. Using a fresh sample, we also successfully repeated the training process above to retain the pink mode and eliminate the blue mode instead.

%We also set up a `test' apparatus where forces can be systematically applied to a set of 7 points along a single sector of this vertex while one of the plates is held clamped.
%Prior to training, forces at 4 of the 7 spots leads to the blue mode while the remaining 3 spots lead to the pink mode.

%The successful training of this simple vertex raises the question of why training was robust despite the relatively sloppy setup of epoxy and imprecise training protocols. For example, visually, we found that no further epoxy was being extruded by the end of training. Despite such an indication that the learning rule had been carried to completion and potentially into overtraining regimes, we found successful training for elimination of a select pathway. To understand the origin of this successful training, we characterized the learned stiffnesses of different creases in the vertex. We cut three samples, trained in an identical manner for elimination of the pink pathway, along three different creases. Each sample was then used to measure the stiffness of the two remaining creases. See SI for details. As shown in Fig.\ref{fig:experiment_setup}C, the change in stiffnesses $\delta \kappa_i$ due to training generally decreases with the (theoretically calculated) folding angle $\rho_i$ at that crease However, the qualitative form of this $\delta \kappa_i(\rho_i)$ relationship resembles the threshold form studied in Fig.\ref{fig:learning_rules}A; in particular, creases have a relatively constant residual stiffness for large enough folding angles. Such residual stiffness, presumably from a minimum amount of epoxy that is retained for any folding angle beyond a threshold, prevents overtraining see in Fig.\ref{fig:learning_rules}C for other learning rules.

\begin{figure}
\centering
\includegraphics[width=0.5\textwidth]{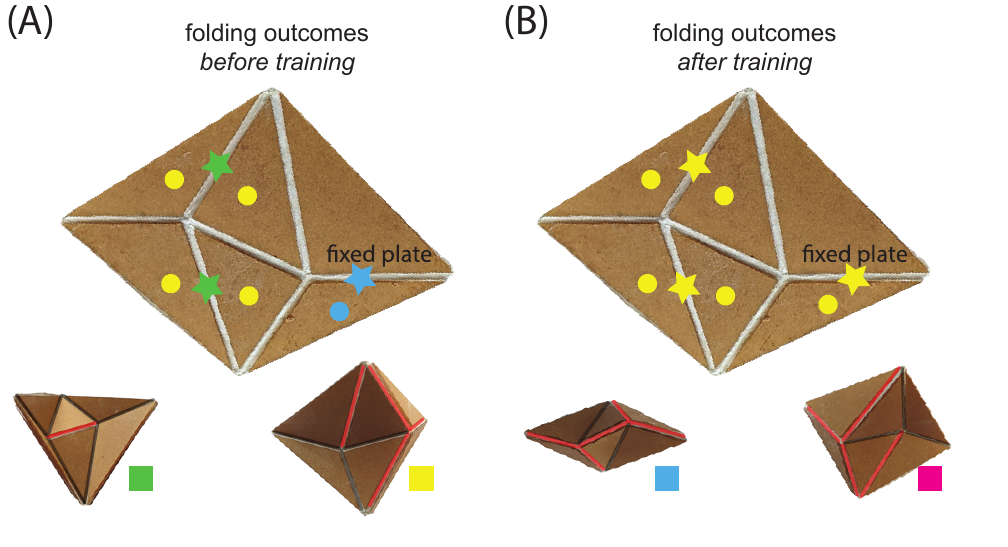}
\caption{Learning to expand a select attractor in a complex sheet.  A sheet with 7 creases and 2 vertices has 4 distinct folding pathways (shown at the bottom). (A) Before training, three pathways are accessible by forces applied to different locations shown (circles) or by torques applied to specific creases (stars). (B) The creases were filled with epoxy and folded back and forth along the yellow pathway as the epoxy set (the training period). After training, the sheet was `tested' with the same forces and torques used in (A). All test forces and torques now result in the yellow pathway, indicating an expanded attractor size for that pathway.} 
\label{fig:Vertex_experiment}
\end{figure}

To see if the principles behind this simple demonstration are robust enough to work in more complex disordered systems, we attempted training on a sheet with 7 creases, 2 vertices and thus 4 distinct pathways at the bifurcation; see Fig.\ref{fig:Vertex_schematic}A. As shown in Fig.\ref{fig:Vertex_experiment}A, the untrained sheet folds along 3 of those 4 pathways for test forces applied to the center of different faces with one face held clamped. (The fourth pathway requires torques at specific creases that cannot be realized by forces at a single face in the clamped configuration we used.) We filled the creases with epoxy and trained with the same protocol as earlier, folding along a select pathway (the yellow pathway), flattening the sheet, folding along the negative branch of that pathway and repeating the process for 60 minutes. After the training process is completed and the epoxy has set, we tested the sheet with the same test forces applied to the faces when it was untrained; we now find that all forces lead to folding along the target yellow pathway as illustrated in Fig.\ref{fig:Vertex_experiment}B. Thus, the flat state bifurcation has been successfully trained to eliminate the other pathways, presumably through saddle-node bifurcations away from the flat state.

%The untrained sheet showed three of the four responses in response to forces applied to the center of different plates with one plate held clamped. The fourth mode (pink in Fig 4C) is not accessible through these limited force patterns and requires special pattern of local torques on creases not easily applied with the clamped set up. We  ..

\section*{Discussion}

The study of bifurcations in mechanical systems has attracted attention from mathematicians~\cite{jordan1999configuration}, roboticists~\cite{wampler1986manipulator,wampler2011mechanism} and physicists~\cite{chen2018branches, stern2017complexity, stern2018shaping}. Most work has focused on changing the structure of bifurcations by rationally changing parameters such as lengths of the elements. Our work here shows at least some versions of this design problem can be solved by changing stiffness of joints through a physical training protocol. Our work further suggests that bifurcations might be physically trainable in mechanical linkages where the lengths of elements change according to learning rules; changes in length have been used as a basis for physical training in other contexts~\cite{pashine2019directed,hexner2019effect,hexner2019periodic,stern2021supervised}. 

The experimental demonstration here illustrates how a generic physical process - the extrusion of material at creases - can naturally implement `learning rules' that confer specific functionality on the system. Other materials naturally show softening with strain~\cite{pashine2019directed}, possibly allowing for the implementation of different functional forms of our learning rules.

The locality of physical learning in mechanical systems contrasts the global nature of most machine learning algorithm, where learning parameters are non-locally updated (gradient descent protocols). The training rules are local, softening each crease in response to strain in that crease. Further, even at the end of training, the learned stiffness in any one crease does not immediately favor one folding pathway over another. However, creases that meet at a vertex~\cite{pinson2017self} have non-linear interactions that constrain their relative folding; such interacting creases are able to collectively learn and encode information about a desired pathway even if each individual crease does not select a pathway by itself.

While we trained for one attractor to grow and occupy most of the force pattern space, one can also train a system for multiple attractors~\cite{stern2020supervised}. Such training can create a mechanical pattern recognizer, folding into one configuration in response to one set of force patterns and a distinct configuration in response to a different set of force patterns. Unlike similar responsive materials designed on a computer, the learning paradigm here lets structures learn \emph{in situ} from real examples of force patterns~\cite{pashine2019directed}.
%Such collective encoding can be contrasted with the more familiar and mechanistically unrelated process of crumpling a sheet of paper; each crease individually learns a rest angle that favors  (and does not eliminate the bifurcation at the flat state.)
%The particular problem here is misleadingly simple - after all, a sheet of paper `remembers' how it was folded. But there are several critical differences. Our system uses the cooperative properties.
% Linear vs non-linear behaviors

\acknowledgements
We thank Heinrich Jaeger, Nidhi Pashine, Chloe Lindeman, Savannah Gowen, and  Martin Falk for discussions. This work was primarily supported by the University of Chicago Materials Research Science and Engineering Center, which is funded by National Science Foundation under award number DMR-2011854. M. S. would like to acknowledge funding from the National Science Foundation via grant DMR-2005749.

\section{Supplementary Information}

\subsection{Theory}

\subsection*{Theoretical Modeling of Self-Folding Sheets}

We model creased sheets using energy-based models developed in previous work  ~\cite{hull2002modelling,tachi2010geometric,pinson2017self}. We assume that creases have a stiffness modeled by torsional spring elements on each crease~\cite{stern2017complexity,stern2018shaping,stern2020supervised}. We briefly review the elements of this model as we build upon this work to simulate the physical learning of desired folding pathways of self folding origami sheets.

% Eqs.~(1-3) in the main text outline how the origami energy is constructed.

The energy of thin sheet origami is dominated by face bending governed by mechanical constraints at the origami vertices. Each vertex contributes $3$ constraints on the folding angles of creases around it~\cite{tachi2010geometric}. Take a vertex surrounded by $N$ creases denoted with an index $i$ and each folded to an angle $\rho_i$. At the flat state, all $\rho_i=0$, which trivially satisfies all mechanical constraints. One can write down an expansion for these $3$ non-linear constraints $T_a(\rho_i)$~\cite{stern2017complexity}. The energy of the folded origami is taken as the sum of squares of the residues of these constraints $E_{\text{Vertex}} \sim \sum_a T_a^2$, which is independently summed over different vertices~\cite{stern2018shaping}. The energy due to the stiffness in the creases is $E_{\text{Crease},i}=\frac{1}{2}\kappa_i\rho_i^2$ as discussed in the main text. Thus, the total energy of a folded sheet is the sum of vertex and stiff crease energies

\begin{equation}
\begin{aligned}
E_{\text{sheet}}(\rho_i) = \sum_{v\in vertices} \sum_{a=1}^3 T_{va}^2 + \frac{1}{2}\sum_{i\in \text{creases}}\kappa_i\rho_i^2.
\end{aligned}
\label{eqn:EnergyModelSI}
\end{equation}

In the learning protocol presented in this work, the crease geometry is fixed and so are the vertex constraints $T_{va}$. The change in the energy of a folded configuration $\rho_i$ during training stems directly from the change in the individual crease stiffness values

\begin{equation}
\begin{aligned}
\frac{dE}{dt}=\frac{\partial E}{\partial \kappa_i} \frac{d\kappa_i}{dt}=\frac{1}{2}\rho_i^2\frac{d\kappa_i}{dt}.
\end{aligned}
\label{eqn:SIEq2}
\end{equation}
%The scale of creases stiffness is denoted by $\bar{k}.$ The choice of stiffness energy scale plays an important role in our learning protocol. We have previously shown how the face bending energy scales like $\rho^4$~\cite{stern2018shaping}, while the crease stiffness energy scales like $\bar{k}\rho^2$. In turn, this gives rise to a transition folding angle scale in our model $\rho_c=\sqrt{\bar{k}}$. For large folding angles $\rho\gg \rho_c$, sheet bending energy dominates, and the folding landscape is controlled solely by the sheet geometry. At small folding angles $\rho \ll \rho_c$ (close to the flat state), crease stiffness dominates, and it is possible to reshape the force-folding map. The goal of training is to reshape this map close to the flat state, such that the applied forces fold the sheet into desired folded states. Throughout this work, we choose an initial uniform crease stiffness $k_i=0.02$. We find that trained sheets, though having heterogeneous stiffness profiles, still maintain a dominant stiffness scale at $\bar{k}\sim 0.02$. In our sheets the transition scale is thus given by $\rho_c\sim\sqrt{0.02}\sim 0.14 rad$, a reasonable angle scale close to the flat state. To make learning in sheets feasible, we conclude that a stiffness scale $\bar{k}\sim 10^{-2}$ should be chosen.

\subsection*{Simulated Sheet Folding}

Using the energy model described previously, we simulated the folding of the self folding origami via several numerical folding methods~\cite{stern2018shaping}. 

%One way that an origami sheet can be folded is by applying torques directly to the different creases. Suppose a crease $i$ of a flat sheet is subjected to an external torque $F^{ext}_i$. Such a torque will induce folding in the crease, but the sheet generally resists folding due to the extra energy that might be associated with a folded structure. Assuming that the folding process is over-damped, we may write a dynamical folding equation

%\begin{equation}
%\tau_{\text{relax}}\frac{d \rho_i}{dt} = - \frac{\partial E_{\text{sheet}}(\bm{\rho})}{\partial \rho_i}  + F_i^{\text{ext}}
%\label{eq:SIFolding},
%\end{equation}

%where $\bm{\rho}$ is the current folded structure, and $\tau_{\text{relax}}$ a time scale of the over-damped dynamics. 

Origami sheets are numerically folded similarly to the way described in~\cite{stern2020supervised}. The folded sheet's configurations are initialized at $$\rho_{i0}=\rho\frac{\tau_i}{\vert\vert\tau_i\vert\vert}$$ by a set of external folding torques ($\tau_i$) on the creases with $\rho\equiv\vert\vert \rho_i\vert\vert$ the folding magnitude, chosen to be $\rho=0.5$. However, this initialization point is typically not an energy minimum on the surface of the hyper-sphere of radius $\rho$. Thus, we numerically relax the sheet to a local minimum of Eq.\ref{eqn:EnergyModelSI} using MATLAB, subject to a constraint that fixes the folding magnitude $\rho$:

\begin{align}
\begin{aligned}
& \underset{\rho_i}{\text{minimize}}
& & E_{\text{sheet}}(\rho_i)\\
& \text{subject to}
& & \vert\vert \rho_i\vert\vert = \rho .
\end{aligned}
\label{eq:SImin}
\end{align}

This protocol mimics the experimental fast folding of origami sheets, and the clamping of one crease at a specific folded dihedral angle. It was tested and validated in~\cite{stern2018shaping}. The results of this folding protocol are similar to torque based folding of sheets using Newtonian methods.

\subsection*{Simulation of Sculpting Folding Pathways Through Physical Learning}

Starting with Eq.\ref{eqn:SIEq2} above we see that the evolution of the energy landscape and, thus, folding pathway is driven by $\frac{d\kappa_i}{dt}$. Physical learning is introduced by the dynamic specification of $\frac{d\kappa_i}{dt}$ defined by various physical learning rules, which are functions of the fold angles of the mode $\rho^{Teacher}_i$, whose attractor size we desire to expand:
\begin{equation} 
\begin{aligned}
d \kappa_i/dt = -\lambda f(|\rho^{Teacher}_i|) \kappa_i.
\end{aligned}
\label{eq:learning_rule_2}
\end{equation}
where $\lambda$, the `learning rate', sets the learning timescale and $\rho^{Teacher}_i$ is a vector defining the fold angles of the creases of the desired mode whose attractor size we want to increase. Note that $\rho^{Teacher}_i$ is obtained by folding the creases of the origami with an external torque, $F^{Teacher}_i$ as described above. We generated the components of $F^{Teacher}_i$ by first randomly selecting a number from a normal distribution. Next we normalized this vector and used it to fold the creases of the origami as described above. We checked if the scalar product between the normalized vector and the resulting normalized folded mode is greater than 0.99. If it is not, we generate another $F^{Teacher}_i$ by selecting another random set of numbers for its components and check if the new vector and its resulting folded mode have a scalar product greater than 0.99. If the scalar product is greater than 0.99, then the final resulting folded mode is then normalized and assigned to $F^{Teacher}_i$. The learning rule is specified by $f(\rho^{Teacher}_i)$ which can be a linear, quadratic or a threshold function of $\rho^{Teacher}_i$. We simulated a physical learning process in which the creases initially had a uniform stiffness $\kappa_i^0=0.02$ (a unit of stiffness represents the bending modulus of the stiff faces of the origami) and evolved with a learning rate $\lambda=0.01$ per training round. Thus, we have specified not just a teacher for the physical learning process, $F^{Teacher}_i$, but also a curriculum or learning rule, $f(\rho^{Teacher}_i)$, for the physical learning process. 

\subsection*{Simulation of Testing Protocol}
After each round of training via physical learning as described above, the attractor size of the desired mode is computed. To calculate the attractor size, a set of an array of test torques, $Test Torque Set$ is created. Each element of the set, $Test Torque Set$ is defined by a vector $F^{external}_i$, whose components, $F^{external}_{i,j}$, represent the magnitude of the torque applied to each crease. Each vector $F^{external}_i$ is normalized and used to fold the creases of the self folding origami sheet as described above. Folding with each external torque, $F^{External}_i$, results in a folded mode $\rho^{Folded}_i$. Note that the folded mode vector, $\rho^{Folded}_i$ is also normalized as well. This folded mode, $\rho^{Folded}_i$ is compared to the desired mode, $\rho^{Teacher}_i$ whose attractor size we seek to increase.

To compare the folded mode, $\rho^{Folded}_i$, with the desired mode, $\rho^{Teacher}_i$, we take the scalar product between both vectors. If the scalar product between the two vectors is above 0.9, then we consider the two modes as similar, and one and the same. We count the number of external torques, $F^{External}_i$, in the set $Test Torque Set$, whose folded mode, $\rho^{Folded}_i$, are considered similar to $\rho^{Teacher}_i$. We then express this count number as a fraction of the cardinality of the set $Test Torque Set$. This fraction defines the attractor size of the desired mode.

\subsection*{Quality of Learning}
Two agents drive the physical learning process: the teacher, $F^{Teacher}_i$, and the curriculum/learning rule, $f(\rho^{Teacher}_i)$. We found the kind of teacher selected does not affect the quality of the physical learning as long as it results in the desired folded mode. Hence, the quality of learning is determined by the kind of learning rule selected. We quantify the quality of learning for various learning rules with two parameters: the peak attractor size attainable and the time robustness of the learning rule. The peak attractor size compares the maximum attractor size achieved for a desired mode for the different learning rules. Meanwhile, the time robustness measures the percentage of the training round for which the self folding origami is optimally trained (i.e. attractor size is above 0.80). The time robustness is a measure of the training protocol's resilience against over-training. 

\subsection*{Calculation of Energy Landscape for A Single Vertex Origami}
To illustrate the mechanism by which physical learning alters the energy landscape of the self folding origami via a saddle-node bifurcation, we plotted the energy, Eq.\ref{eqn:EnergyModelSI} of the different folded configurations at several points during the training of the origami. After each round of training, the stiffnesses in the creases of the single vertex changes and the energy landscape of the folded configuration space is re-computed.This time-energy landscape plot shows the elimination of the unwanted folding pathway (mode) via a saddle-node bifurcation, and the preservation of the desired folding pathway (mode) after several rounds of training. Further training results in a recovery of the previously eliminated mode. 

\subsection{Experiments}

\subsection*{Acrylic sheet setup}

To create a system naturally capable of learning, we exploited an origami system with fresh epoxy totally filled into the crease pattern of the origami. This epoxy is extruded from the creases during the folding of the origami during the training protocol. This results in a final stiffness (after the epoxy sets) that depends on the amount of folding of each of the creases.
We laser cut origami patterns in acrylic sheets of thickness 1.5mm; crease lines were designed to have a gap (or width) of 30mm. Test holes of 10mm diameter are laser cut on the acrylic sheet at various strategic positions (along the circumference of the 90$^{\circ}$ plate for the single vertex and on the center of each plate of the two-vertex). Two copies of such acrylic patterns were each glued to both sides of a sheet of Tyvek.% using Duco Cement. 
The corresponding plates of the origami patterns on each side of the sheet of Tyvek are lined up with each other before the glue sets. After the glue is set, holes for applying testing forces on the faces of the origami plates are perforated. The resulting setup has stiff faces (bending stiffness set by acrylic) and soft creases (stiffness set by the Tyvek sheet). Origami patterns studied were for a single-vertex and two-vertex. The single vertex has four creases radiating from a single vertex at the center of the pattern. The creases of this single vertex pattern form sector-angles 150$^{\circ}$, 60$^{\circ}$, 90$^{\circ}$, and 60$^{\circ}$. The two-vertex pattern with a total of 7 creases consists of two internal vertices; one vertex has the following sector angles 107$^{\circ}$, 123$^{\circ}$, 82$^{\circ}$, and 48$^{\circ}$, while the second vertex is surrounded by sector angles 82$^{\circ}$, 54$^{\circ}$, 99$^{\circ}$, and 125$^{\circ}$. The two vertices are connected by a common crease. This connecting crease serves as the boundary dividing the 82$^{\circ}$ sector plate from the 48$^{\circ}$ sector plate of the first vertex, and the 82$^{\circ}$ sector plate from the 54$^{\circ}$ sector plate of the second vertex.  The sector angles, crease lengths, and position of the vertices for both the single-vertex and two-vertex are specified in a supplementay PDF file which can be used to laser cut these patterns.
%To quantitatively characterize the learning rule (Figure 3), we created a simple assembly with a single crease of width 30mm. This is simply two pairs of rectangular sheets of acrylic (measuring 90mm by 50mm) glued on both sides of the Tryvek sheet and separated by a crease of width 30mm. The corresponding rectangular acrylic plate glued on both sides of the Tryvek sheet are lined up with each other.

\subsection*{Epoxy and training}

\paragraph{Epoxy Mixture}

An epoxy solution is made by mixing epoxy resin with its curing agent in the ratio 1:2. This mixture is stirred for about five minutes and poured into the creases of the origami pattern on both sides of the assembly. Note, that if the epoxy had been mixed in the ratio 1:1, upon curing, it would be so stiff that the origami assembly would be difficult to fold, without destroying the assembly; such epoxy when hardened is also brittle and would fracture under a bending moment attempting to fold the origami assembly. Hence, we mix epoxy resin and curing agent in the ratio 1:2, allowing for crease folding upon curing of the epoxy mixture, without disintegration of the origami assembly.

\paragraph{Folding of Origami Assembly}
The origami assembly filled with watery epoxy in its creases is folded into the desired configuration that is to be trained for. The origami assembly is trained by folding the assembly back and forth along the positive and negative components of a desired folding pathway. This cyclic folding between both pairs of the desired mode is repeated for an hour, during which the epoxy solution begins to cure and is no longer watery. A simple folding protocol is utilized to fold the origami assembly into the desired configuration: one of the plates of the origami assembly is fixed while pushing or pulling on any of the other plates of the origami assembly with a normal force or a turning torque exerted at a single contact point on any of the non-fixed origami plates.

\paragraph{Training Under Gravity}
Since the epoxy is still watery during origami training, it needs to be trained on a rotating platform to avoid epoxy from flowing out of the creases due to the influence of gravity. The rotating platform consists of two standing laboratory clamps screwed to the optical table and situated 600mm apart.  A rod is horizontally supported by the claws of the two standing laboratory clamps, but the rod is allowed to freely rotate within the claws of the clamp. A lab clamp retort is then fixed clamped (allows for no rotation or slipping) to the rotation-free horizontally placed rod with the claws on one end of lab clamp retort, while the claws on the other end of the lab clamp retort is clamped to one of the plates of the origami (the plate fixed during training as previously described).
As one hand is used to exert a normal force or turning torque on one of the free plates of the origami assembly in order to fold it into the positive and negative components of the desired mode, the other hand is used to manually turn the rotation-free horizontal rod. This combined process results in the folding of the origami assembly, while under rotation, and, thus, prevents the flow of the epoxy solution from the origami’s creases during this training; while under the influence of gravity.

\subsection*{Emergence of Crease Stiffness}
After training the origami assembly, the origami samples are allowed to further cure and are left hanging in the lab for a week, allowing the epoxy solution in the creases to harden, thus, producing an effective stiffness on the creases of the origami.

\subsection*{Testing protocol}
The plate of the origami assembly fixed during training is clamped. A 100mm long thread knotted on one end is passed through each of the testing holes of the origami assembly. The threads are gently pulled normal to the surface of the origami plate. Upon pulling each thread in each hole, the origami folds into a given configuration. The resulting configuration for each pull is recorded. The attractor size of the different folding modes of the origami assembly is computed. This process is repeated for both the trained and untrained samples of the origami assembly. The attractor size of the chosen trained mode before and after training are compared to one another.

%\bibliography{Bibliography}
%\bibliographystyle{unsrt}

\end{document}